\documentclass[12pt]{article}

\usepackage{amsmath}
\usepackage{amssymb}
\begin{document}

\title{Bogomolny equations for the BPS Skyrme models with impurity}
\author{{\L}. T. St\c{e}pie\'{n} \thanks{The Pedagogical University of Cracow, ul. Podchora\c{}\.{z}ych 2, 30-084 Krak\'{o}w, Poland; 
sfstepie@cyf-kr.edu.pl}}
\vspace{10pt}

\maketitle

\begin{abstract}
We show that the BPS Skyrme model as well as its (2+1) dimensional baby version can be coupled with an impurity in the BPS preserving manner. The corresponding Bogomolny equations are derived. 
\end{abstract}



\section{Introduction}

Addition of an impurity i.e., a non-dynamical background field, physically represents a non-trivial medium, in which a given field theory is immersed. Such medium effects may be due to local 
non-homogeneities as well as global dependence on external parameters. Especially, it is of some relevance for realistic solitonic systems, where typically a soliton exists not as an isolated object surrounded by vacuum, but, on the contrary, it exists in a presence of other solitons. Hence, an impurity can be also regarded as an averaged (integrated) impact of other solitons or as a frozen set of solitons (spectators). 

A very important class of solitonic models are the so-called BPS theories \cite{SM}, \cite{Shnir}. Such a model possesses topologically nontrivial solutions saturating a pertinent topological lower bound on the energy. Because of that, the solutions are stable and follow from so-called Bogomolny equations, which are of lower order than the full Lagrange-Euler equations. Therefore, these models are much easier to solve than a generic model, and provide us a very important mathematical insight, especially in the case of higher dimensional topological solitons. Here let us mention the Abelian Higgs model at the critical coupling, 't Hooft-Polyakov monopole or Yang-Mills instantons as the most prominent examples. BPS property indicates for a given model that there is possible to derive 
Bogomolny equations for this model - such equations were derived for restricted BPS baby Skyrme model, \cite{AdamEtAl2}, \cite{LS1}, 
for gauged restricted BPS baby Skyrme model, \cite{LS2}. 
Some aspects of relations between supersymmetry and Bogomolny equations are considered in \cite{BogomolnySupersymmetry}.  
In \cite{arxiv1406.7647}, \cite{arxiv1504.08123}, higher derivative field-theoretical models (among others, generalized Skyrme models) and BPS states especially 
 in the supersymmetric setups, were investigated, and various Bogomolny equations for these models were derived. 
 Next, in \cite{arxiv1408.4210} higher derivative corrections to supersymmetric realizations in the off-shell superfield formalism
 (without the problem of the auxiliary field) were constructed. A supersymmetric extension of the Skyrme term without the problem of the auxiliary field,
 was constructed in \cite{arxiv1512.07557}. Some solitons and instantons (non-BPS), for supersymmetric Skyrme model, were found
 in \cite{arxiv1608.03526} (the first pure Skyrme instanton had been presented in \cite{Speight_pureSkyrmeInstant}).

Usually, coupling of an impurity to an arbitrary BPS model,  breaks the BPS property completely, see for example \cite{imp1}-\cite{imp5}. Physically it means that static solitons, which in the original BPS theory do not interact with each other, begin to feel an attractive or repulsive force due to the presence of the impurity. In a consequence, the moduli space of energetically equivalent solutions (in the fixed topological sector) disappears as the solitons have energetically preferred positions (basically fixing the mutual distance). This has a nontrivial impact on solvability of the model (no Bogomolny equations) as well as its static and dynamical properties. Some vortex solutions in presence of magnetic impurity, were found in \cite{vortices_impurities}. 

It has been recently found that there it is possible to couple an impurity to a BPS model, in such a way that half of the BPS property remains preserved \cite{BPS_impurity1}, \cite{BPS_impurity2}. It means that half of originally BPS solitons keep this feature also in the presence of impurity \cite{BPS_impurity2}, \cite{BPS_impurity3}. Hence, for these solutions  there is some well defined moduli space. This was originally applied to (1+1) dimensional models with kinks and to some (2+1) planar solitons models \cite{BPS_impurity4}. 

A theory, where inclusion of impurities seems to be of great importance, is the Skyrme model \cite{skyrme}. This is one of the most popular effective theories of the low energy QCD describing non-perturbative excitations i.e., baryons and atomic nuclei as coherent solutions in the "mesonic fluid", that is (semiclassically quantized) solitons in a theory with pionic degrees of freedom \cite{nappi}, \cite{wood}, \cite{manton} (fields in the Lagrangian). Furthermore, the Skyrme model is considered as relevant for description of nuclear matter at high density, where in-medium effects must nontrivially deform the theory. For our purposes it is important to note that at high density regime (relevant for in-medium/impurities effects), the generalized Skyrme model approaches a BPS theory, known as the BPS Skyrme model \cite{BPS}. This limit of the generalised Skyrme model is considered to be crucial for description of (the inner core of) neutron stars \cite{NS}, \cite{carlos}. Hence, an obvious question arrises, whether (3+1) dimensional BPS Skyrme model can be coupled with an impurity in a self-duality preserving manner. 
In \cite{Schroers} some class of models for static magnetic skyrmions, with presence of the Dzyaloshinskii-Moriya interaction, is investigated; 
it has been shown, that one can consider impurity as a non-abelian gauge field, and the impurity equation found in \cite{BPS_impurity4}, was derived from Bogomolny equation.

In this paper, we show that the BPS Skyrme model admits a coupling to an impurity such that the BPS property is kept unchanged. In particular we derive the Bogomolny equation for BPS family Skyrme models with a general form of the potential and with the impurity. We do it in a systematic way, namely, we apply the concept of strong necessary conditions (CSNC) presented in \cite{S1}, 
and developed in \cite{SWL1}, \cite{SWL2}, \cite{SWL3}, \cite{AdamSantamaria}. The application of the concept of
of CSNC, for deriving Bogomolny decomposition, was included in \cite{SSS} (there were also described the relations between supersymmetry and Bogomolny decomposition derived by using CSNC).  

This paper is organized, as follows. The next section contains a short description of the concept of strong necessary conditions. In the section 3, we derive Bogomolny equation for BPS family Skyrme models, by using this concept. In the section 4, we sum up the results, obtained in this paper.

   \section{The concept of strong necessary conditions - a brief description}
   
The derivation of Bogomolny equations for some field theory systems in lower dimension, by using the CSNC method, was presented in \cite{SSS}, \cite{S},  \cite{SSS2}. It was also applied to baby Skyrme theories and their gauged version, in \cite{LS1} and \cite{LS2}, correspondingly. 

Here we briefly describe this method. As this is well-known fact, the Euler-Lagrange equations, 
 
 \begin{equation}
 F_{,u} - \frac{d}{dx}F_{,u_{,x}} - \frac{d}{dy}F_{,u_{,y}}=0, \label{el}
 \end{equation}

  follow from $\delta \Phi[u]$, where $\Phi[u]$ is the functional
 
 \begin{equation}
 \Phi[u]=\int_{E^{2}} F(u,u_{,x},u_{,y}) \hspace{0.05 in} dxdy. \label{functional}
 \end{equation}

 Instead of considering of (\ref{el}), we consider strong necessary conditions, \cite{S1}, \cite{SWL1}, \cite{SWL2}

 \begin{gather}
   F_{,u}=0, \label{silne1} \\
   F_{,u_{,x}}=0, \label{silne2} \\
   F_{,u_{,y}}=0, \label{silne3}
 \end{gather} 

 where $F_{,u} \equiv \frac{\partial F}{\partial u}$, etc.
   
 Of course, the set of all solutions of the system of the equations (\ref{silne1}) - (\ref{silne3}),
 is a subset of the set of solutions of the Euler-Lagrange equation (\ref{el}). 
 However, if there exist the solutions  of the equations (\ref{silne1}) - (\ref{silne3}), 
 they are very often trivial solutions. So, in order to avoid this triviality, 
 one can make gauge transformation of the functional (\ref{functional})
 
  \begin{equation}
  \Phi \rightarrow \Phi + Inv, \label{gauge_transf}
  \end{equation}

  where $Inv$ is such functional that its local variation vanishes, with respect to $u(x,y)$:
 $\delta Inv \equiv 0$.
 
  Owing to this feature, the Euler-Lagrange equations (\ref{el}) and the Euler-Lagrange equations resulting from  requiring of the extremum of $\Phi + Inv$, are equivalent.
 On the other hand, the strong necessary conditions (\ref{silne1}) - (\ref{silne3}) are not invariant with respect to the gauge transformation (\ref{gauge_transf}). 
 Hence, now is a chance to obtain non-trivial solutions. Let us note that the order of the system 
 of the partial differential equations, following from strong necessary conditions 
 (\ref{silne1}) - (\ref{silne3}), is less than the order of Euler-Lagrange equations (\ref{el}). 

 \section{The derivation of the BPS equation for two models with impurity}

Now we show, how to apply the concept of strong necessary conditions in practice. 
We derive the Bogomolny equation for both models with impurity: BPS Skyrme model and restricted 
baby BPS Skyrme model. 

 \subsection{The case of restricted baby BPS Skyrme model with impurity}

 According to the idea presented in the previous section, we make the gauge transformation of 
 the density of the energy functional of the restricted baby BPS Skyrme model: $\mathcal{H} \longrightarrow \mathcal{\tilde{H}}$, however, in order to generalize the investigations, we consider

 \begin{equation}
 \begin{gathered}
   H = \int dx^{2} \mathcal{H} = \int d^{2}x (f \cdot (\varepsilon^{mn} i \omega_{,m} \omega^{\ast}_{,n})^{2} + V + \\
	\lambda_{1} \sigma(x^{k}) (\varepsilon^{mn} i \omega_{,m} \omega^{\ast}_{,n}) + \lambda_{2} \sigma(x^{k}) \sqrt{V} 
	+ \sigma^{2}(x^{k})),  
	\label{hamiltonian}
  \end{gathered}
  \end{equation}
	
   where $f, V \in \mathcal{C}^{1}$ are some unspecified functions of $\omega, \omega^{\ast}$ and $k, m, n = 1, 2$.
  (For the sake of generality, we leave $f$ as a completely unspecified function, which correspond to a general target space geometry.) 
	Next, we have to establish 
  the general form of the density of the topological invariant $I_{1}=\int d^{2}x \mathcal{I}_{1}$, i.e. $\delta I_{1} \equiv 0$. 
  It turns out that this density has the following form:
  \begin{gather}
  \mathcal{I}_{1} = G_{1} \varepsilon^{mn} \omega_{,m} \omega^{\ast}_{,n}, \label{glowny_niezm}
  \end{gather}

  where $G_{1} \in \mathcal{C}^{1}$ is some arbitrary function of $\omega,\omega^{\ast}$.\\

  The density of the gauged energy functional has the following form:
	
	\begin{equation}
  \begin{gathered}
  \mathcal{\tilde{H}} = f \cdot (\varepsilon^{mn} i \omega_{,m} \omega^{\ast}_{,n})^{2} + V + \\
	\lambda_{1} \sigma(x^{k}) (\varepsilon^{mn} i \omega_{,m} \omega^{\ast}_{,n}) + \lambda_{2} \sigma(x^{k}) \sqrt{V} 
	+ \sigma^{2}(x^{k}) + \sum^{3}_{r=1} \mathcal{I}_{r}, 
  \end{gathered}
	\end{equation} 

where  $f=f(\omega, \omega^{\ast}), V = V(\omega, \omega^{\ast}) \in \mathcal{C}^{1}, k, m, n = 1, 2$. $\mathcal{I}_{1}$ is given by (\ref{glowny_niezm}), and $\mathcal{I}_{p+1}$ are the densities of so-called divergent invariants: $\mathcal{I}_{p+1} = D_{p}G_{p+1}, 
 p=1, 2$. $G_{1}$ and $G_{p+1} \in \mathcal{C}^{1}$, are some functions of $\omega, \omega^{\ast}$, to be determined later.\\
 The applying of strong necessary conditions, gives the so-called dual equations:
  \begin{gather}
  \mathcal{\tilde{H}}_{,\omega} : f_{,\omega} \cdot  (\varepsilon^{mn} i \omega_{,m} \omega^{\ast}_{,n})^{2} + V_{,\omega} + 
  \lambda_{2} \frac{V_{,\omega}}{2\sqrt{V}} \sigma +  
  G_{1,\omega} \varepsilon^{mn} \omega_{,m} \omega^{\ast}_{,n} + \sum^{2}_{p=1} D_{p} G_{p+1,\omega} = 0, \label{gorne1} \\
  \mathcal{\tilde{H}}_{,\omega^{\ast}} : f_{,\omega^{\ast}} \cdot  (\varepsilon^{mn} i \omega_{,m} \omega^{\ast}_{,n})^{2} + V_{,\omega^{\ast}} +  \lambda_{2} \frac{V_{,\omega^{\ast}}}{2\sqrt{V}} \sigma + 
  G_{1,\omega^{\ast}} \varepsilon^{mn} \omega_{,m} \omega^{\ast}_{,n} + \sum^{2}_{p=1} D_{p} G_{p+1,\omega^{\ast}} = 0, \label{gorne2} \\
  \mathcal{\tilde{H}}_{,\omega_{,r}} : 2 f \cdot \varepsilon^{rj} i \omega^{\ast}_{,j}  (\varepsilon^{mn} i  \omega_{,m}  
  \omega^{\ast}_{,n}) + \lambda_{1} \sigma \varepsilon^{rj} i \omega^{\ast}_{,j} + G_{1} \varepsilon^{rj}  \omega^{\ast}_{,j} + G_{r+1,\omega} = 0, \label{dolne1}\\
  \mathcal{\tilde{H}}_{,\omega^{\ast}_{,r}} : 2 f \cdot \varepsilon^{jr} i \omega_{,j}  (\varepsilon^{mn}  i \omega_{,m}  
  \omega^{\ast}_{,n}) + \lambda_{1} \sigma \varepsilon^{jr} i \omega_{,j} + G_{1} \varepsilon^{jr}  \omega_{,j} + G_{r+1,\omega^{\ast}} = 0. \label{dolne2}
  \end{gather} 
Now we need to make the equations (\ref{gorne1}) - (\ref{dolne2}) self-consistent. At first, we want  the equations (\ref{dolne1}) - (\ref{dolne2}) to be self-consistent.
  In this order, we put

   \begin{equation}
   \begin{gathered}
   2 i f \cdot  (\varepsilon^{mn} i \omega_{,m} \omega^{\ast}_{,n}) + i \lambda_{1} \sigma + G_{1} = 0, \label{uzg} \\
   G_{r+1}=const, r=1,2.
   \end{gathered}
   \end{equation}    
 Hence, instead of two equations (\ref{dolne1}) - (\ref{dolne2}), we have one equation:
   \begin{gather}
   2 i f \cdot  (\varepsilon^{mn} i \omega_{,m} \omega^{\ast}_{,n}) + i \lambda_{1} \sigma + G_{1} = 0,  \label{Bogomolny_baby}
   \end{gather}
where $f=f(\omega, \omega^{\ast}) \in \mathcal{C}^{1}, G_{1}=G_{1}(\omega, \omega^{\ast}) \in \mathcal{C}^{1}$.

Now we have to eliminate from 
   (\ref{gorne1}) - (\ref{gorne2}), all terms including the derivatives of the fields. In this order we use (\ref{uzg}). Hence, we have two equations for the function $G_{1}$ and for the potential $V(\omega, \omega^{\ast})$

   \begin{equation}
   \begin{gathered}
   -\frac{1}{4} \frac{f_{,\omega} (i \lambda_{1} \sigma+ G_{1})^{2}}{f^{2}} + V_{,\omega} + 
	\frac{\lambda_{1} \sigma V_{,\omega}}{2\sqrt{V}} +  	
	\frac{1}{2}  \frac{G_{1,\omega} (i \lambda_{1} \sigma+ G_{1})}{f} = 0,\\
   -\frac{1}{4} \frac{f_{,\omega^{\ast}} (i \lambda_{1} \sigma+ G_{1})^{2}}{f^{2}} + V_{,\omega^{\ast}} + 
	 \frac{\lambda_{1} \sigma V_{,\omega^{\ast}}}{2\sqrt{V}} +  
	\frac{1}{2}  \frac{G_{1,\omega^{\ast}} (i \lambda_{1} \sigma+ G_{1}) G_{1}}{f} = 0.
   \end{gathered}
   \end{equation}
  Their solution is:

   \begin{gather}
  G_{1} = -i\lambda_{1} \sigma -  \sqrt{f (-4 \lambda_{1} \sigma \sqrt{V} - 4 V + C_{1})} . \label{funkcjaG1}
   \end{gather}	
	
	 So we put $C_{1}=-C_{2}$:
	
	  \begin{gather}
  G_{1} = -i\lambda_{1} \sigma - i \sqrt{f (4 \lambda_{1} \sigma \sqrt{V} + 4 V + C_{2})} . \label{funkcjaG_1}
   \end{gather}

	and we can eliminate $G_1$ from (\ref{Bogomolny_baby}). Hence, we have the Bogomolny equation for the  restricted baby BPS Skyrme model with impurity
	
	 \begin{gather}
   2 f \cdot  (\varepsilon^{mn} i \omega_{,m} \omega^{\ast}_{,n})   =  \sqrt{f (4 \lambda_{1} \sigma \sqrt{V} + 4 V + C_{2})}. \label{Bogomolny_baby_final}
   \end{gather}

 \subsection{The case of BPS Skyrme model with impurity}

 BPS Skyrme model with impurity is a mathematical generalization of the (restricted) baby BPS Skyrme model with impurity. 
 The essence of this generalization is such that there in the hamiltonian appears the additional scalar field $\chi$, so 
 the functions $f, V \in \mathcal{C}^{1}$ are now some unspecified functions of $\chi, \omega, \omega^{\ast}$, and 
 each such function depends on three independent variables: $x, y, z$.

 Again, we proceed according to the idea of the concept of strong necessary conditions. Namely, we make the gauge transformation of the density of the energy functional of the BPS Skyrme model: $\mathcal{H} \longrightarrow \mathcal{\tilde{H}}$, however, in order to generalize the investigations, we consider

  \begin{equation}
  \begin{gathered}
   H = \int d^{3} x \mathcal{H} = \int d^{3} x (f \cdot (\varepsilon^{kmn} i \chi_{,k} \omega_{,m} \omega^{\ast}_{,n})^{2} + V + \\
	\lambda_{1} \sigma(x^{l}) (\varepsilon^{kmn} i \chi_{,k} \omega_{,m} \omega^{\ast}_{,n}) + \lambda_{2} \sigma(x^{l}) \sqrt{V} 
	+ \sigma^{2}(x^{k})),  
	\label{hamiltonianBPS}
  \end{gathered}
	\end{equation} 
	
   where $f, V \in \mathcal{C}^{1}$ are some unspecified functions of $\chi, \omega, \omega^{\ast}$ and $k, l, m, n = 1, 2, 3$ (as in the previous model, for the sake of generality, we assume that $f$ is some completely unspecified function, corresponding to a general target space geometry. The usual $S^3$ target space is obtained for $f=\frac{\sin^4 \chi}{(1+|\omega|^2)^4}$.) Next, we have to establish the general form of the density of the topological invariant $I_{1}=\int d^{3}x \mathcal{I}_{1}$, i.e. $\delta I_{1} \equiv 0$. It turns out that this 
  density has the following form:
  \begin{gather}
  \mathcal{I}_{1} = G_{1} \varepsilon^{kmn}\chi_{,k} \omega_{,m} \omega^{\ast}_{,n}, \label{glowny_niezmBPS}
  \end{gather}
where $G_{1} \in \mathcal{C}^{1}$ is some arbitrary function of $\chi, \omega,\omega^{\ast}$.
\\
The density of the gauged energy functional has the following form:

  \begin{equation}
  \begin{gathered}
  \mathcal{\tilde{H}} = f \cdot (\varepsilon^{kmn} i \chi_{,k} \omega_{,m} \omega^{\ast}_{,n})^{2} + V + \\
	\lambda_{1} \sigma(x^{l}) (\varepsilon^{kmn} i \chi_{,k} \omega_{,m} \omega^{\ast}_{,n}) + \lambda_{2} \sigma(x^{l}) \sqrt{V} 
	+ \sigma^{2}(x^{l}) + \sum^{4}_{r=1} \mathcal{I}_{r},
  \end{gathered}
	\end{equation}
	
where  $f=f(\chi, \omega, \omega^{\ast}), V = V(\chi, \omega, \omega^{\ast}) \in \mathcal{C}^{1}$, $k, l, m, n = 1, 2, 3$. $\mathcal{I}_{1}$ is given by (\ref{glowny_niezm}), and $\mathcal{I}_{p+1}$ are the densities of so-called divergent invariants: $\mathcal{I}_{p+1} = D_{p}G_{p+1}, 
 p=1,...,3$. $G_{1}$, and $G_{p+1} \in \mathcal{C}^{1}$, are some functions of $\chi, \omega, \omega^{\ast}$, to be determined later.\\
 The applying of strong necessary conditions, gives the so-called dual equations:

  \begin{gather}
  \mathcal{\tilde{H}}_{,\chi} : f_{,\chi} \cdot  (\varepsilon^{kmn} i \chi_{,k} \omega_{,m} \omega^{\ast}_{,n})^{2} + V_{,\chi} + 
  \lambda_{2} \sigma \frac{V_{,\chi}}{2 \sqrt{V}} +  
  G_{1,\chi} \varepsilon^{kmn}\chi_{,k} \omega_{,m} \omega^{\ast}_{,n} + \sum^{3}_{p=1} D_{p} G_{p+1,\chi} = 0, \label{gorne11} \\
  \mathcal{\tilde{H}}_{,\omega} : f_{,\omega} \cdot  (\varepsilon^{kmn} i \chi_{,k} \omega_{,m} \omega^{\ast}_{,n})^{2} + V_{,\omega} + 
  \lambda_{2} \sigma \frac{V_{,\omega}}{2 \sqrt{V}} +  
  G_{1,\omega} \varepsilon^{kmn}\chi_{,k} \omega_{,m} \omega^{\ast}_{,n} + \sum^{3}_{p=1} D_{p} G_{p+1,\omega} = 0, \label{gorne22} \\
  \mathcal{\tilde{H}}_{,\omega^{\ast}} : f_{,\omega^{\ast}} \cdot  (\varepsilon^{kmn} i \chi_{,k} \omega_{,m} \omega^{\ast}_{,n})^{2} + V_{,\omega^{\ast}} +  \lambda_{2} \sigma \frac{V_{,\omega^{\ast}}}{2 \sqrt{V}} + 
  G_{1,\omega^{\ast}} \varepsilon^{kmn}\chi_{,k} \omega_{,m} \omega^{\ast}_{,n} + \sum^{3}_{p=1} D_{p} G_{p+1,\omega^{\ast}} = 0, \label{gorne33} \\
  \mathcal{\tilde{H}}_{,\chi_{,r}} : 2 f \cdot \varepsilon^{rsj} i \ \omega_{,s} \omega^{\ast}_{,j}  (\varepsilon^{kmn} i \chi_{,k} \omega_{,m}  
  \omega^{\ast}_{,n}) + \sigma \varepsilon^{rsj} i \omega_{,s} \omega^{\ast}_{,j} + G_{1} \varepsilon^{rsj} \omega_{,s} \omega^{\ast}_{,j} + G_{r+1,\chi} = 0 , \label{dolne11}\\
  \mathcal{\tilde{H}}_{,\omega_{,r}} : 2 f \cdot \varepsilon^{srj} i \ \chi_{,s} \omega^{\ast}_{,j}  (\varepsilon^{kmn} i \chi_{,k} \omega_{,m}  
  \omega^{\ast}_{,n}) + \sigma \varepsilon^{srj} i \chi_{,s} \omega^{\ast}_{,j} + G_{1} \varepsilon^{srj} \chi_{,s} \omega^{\ast}_{,j} + G_{r+1,\omega} = 0, \label{dolne22}\\
  \mathcal{\tilde{H}}_{,\omega^{\ast}_{,r}} : 2 f \cdot \varepsilon^{sjr} i \ \chi_{,s} \omega_{,j}  (\varepsilon^{kmn} i \chi_{,k} \omega_{,m}  
  \omega^{\ast}_{,n}) + \sigma \varepsilon^{sjr} i \chi_{,s} \omega_{,j} + G_{1} \varepsilon^{sjr} \chi_{,s} \omega_{,j} + G_{r+1,\omega^{\ast}} = 0. \label{dolne33}
  \end{gather} 
Now we need to make the equations (\ref{gorne11}) - (\ref{dolne33}), self-consistent. At first, analogically to the previous model, we want  the equations (\ref{dolne11}) - (\ref{dolne33}) to be self-consistent.
  In this order, we put

   \begin{equation}
   \begin{gathered}
   2 i f \cdot  (\varepsilon^{kmn} i \chi_{,k} \omega_{,m} \omega^{\ast}_{,n}) + i \lambda_{1} \sigma + G_{1} = 0, \label{uzgBPS} \\
   G_{r+1}=const, r=1,2,3.
   \end{gathered}
   \end{equation}    
 Hence, instead of three equations (\ref{dolne11}) - (\ref{dolne33}), we have one equation:
   \begin{gather}
   2 i f \cdot  (\varepsilon^{kmn} i \chi_{,k} \omega_{,m} \omega^{\ast}_{,n}) + i \lambda_{1} \sigma + G_{1} = 0,  \label{Bogomolny}
   \end{gather}
where $f=f(\chi, \omega, \omega^{\ast}) \in \mathcal{C}^{1}, G_{1}=G_{1}(\chi, \omega, \omega^{\ast}) \in \mathcal{C}^{1}$. Next, we again eliminate from 
   (\ref{gorne11}) - (\ref{gorne33}), all terms including the derivatives of the fields, by using (\ref{uzg}). Hence, we have three equations for the function $G_{1}$ and for the potential $V(\chi, \omega, \omega^{\ast})$

   \begin{equation}
   \begin{gathered}
   -\frac{1}{4} \frac{f_{,\chi} (i\lambda_{1} \sigma+ G_{1})^{2}}{f^{2}} +  V_{,\chi} + \frac{\lambda_{1} \sigma V_{,\chi}}{2\sqrt{V}} + 	
	\frac{1}{2}  \frac{G_{1,\chi} (i\lambda_{1} \sigma+ G_{1})}{f} = 0,\\
   -\frac{1}{4} \frac{f_{,\omega} (i\lambda_{1} \sigma+ G_{1})^{2}}{f^{2}} + V_{,\omega} + 
	\frac{\lambda_{1} \sigma V_{,\omega}}{2\sqrt{V}} +  	
	\frac{1}{2}  \frac{G_{1,\omega} (i\lambda_{1} \sigma+ G_{1})}{f} = 0,\\
   -\frac{1}{4} \frac{f_{,\omega^{\ast}} (i\lambda_{1} \sigma+ G_{1})^{2}}{f^{2}} + V_{,\omega^{\ast}} + \frac{\lambda_{1} \sigma V_{,\omega^{\ast}}}{2\sqrt{V}} +  
	\frac{1}{2}  \frac{G_{1,\omega^{\ast}} (i\lambda_{1} \sigma + G_{1})}{f} = 0.
   \end{gathered}
   \end{equation}
	
Their solution is analogical to the function $G_{1}$ in the case of baby BPS Skyrme model with impurity:

   \begin{gather}
  G_{1} = -i\lambda_{1} \sigma - i \sqrt{f (4 \lambda_{1} \sigma \sqrt{V} + 4 V + C_{2})}, \label{funkcja_G1}
   \end{gather}
	
	where instead of the integration constant $C_{1}$, again the constant $C_{2}$ appears ($C_{1}=-C_{2}$).
	
   Thus, we can again eliminate $G_1$ from (\ref{Bogomolny}). Then, the Bogomolny equation for the BPS Skyrme model with impurity
	has the form:
		
   \begin{equation}
      2 f \cdot  (\varepsilon^{kmn} i \chi_{,k} \omega_{,m} \omega^{\ast}_{,n})   = 
			\sqrt{f (4 \lambda_{1} \sigma \sqrt{V} + 4 V + C_{2})}.
   \end{equation}

\section{Summary}

In this paper the Bogomolny equations for the BPS Skyrme model and restricted baby BPS Skyrme model, with impurity terms and with the general forms of the potentials $V$, were derived, by using the concept of strong necessary conditions method.
As one can see, in the absence of the impurity term ($\sigma=0$), the found Bogomolny equations and conditions for the functions
$G_{1}$ become into their versions for BPS Skyrme model, \cite{LS2016} and restricted BPS baby Skyrme model without impurity terms \cite{LS1} 
(in the last case, $C_{2}=0$). 
 
This result provides the first example of BPS preserving impurity model in (3+1) dimension. 

\section{Acknowledgments}
The author thanks to Prof. A. Wereszczy\'{n}ski for his valuable remarks. 

\section{Computational resources} 
  
   Some part of computations was carried out by using Maple Waterloo Software, owing to the financial support, provided by The Pedagogical University of Cracow, within a research project (the leader of this project: Dr K. Rajchel).

\section*{References}

\end{document}